\documentclass[12pt]{article}

\usepackage{amsmath,amssymb,epsfig,epic}

\usepackage{color}
\begin{document}
\title{Searching for integrable lattice maps using factorization}
\author{Jarmo Hietarinta\\ Department of Physics, University of Turku\\ 
FIN--20014 Turku, Finland\\ \\
Claude Viallet \\
UMR  7589 Centre National de la Recherche Scientifique \\
 Universit\'e Paris VI, Universit\'e Paris VII  \\
Bo\^\i te 126,  4 place Jussieu \\ 
F--75252 Paris Cedex 05, France}

\maketitle

\begin{abstract}
We analyze the factorization process for lattice maps, searching for
integrable cases.  The maps were assumed to be at most quadratic in
the dependent variables, and we required minimal factorization (one
linear factor) after 2 steps of iteration. The results were then
classified using algebraic entropy.  Some new models with polynomial
growth (strongly associated with integrability) were found. One of
them is a nonsymmetric generalization of the homogeneous quadratic
maps associated with KdV (modified and Schwarzian), for this new model
we have also verified the ``consistency around a cube''.
\end{abstract}

\section{Introduction}
Although many interesting results have been obtained for discrete integrable
systems and many properties have been clarified, we are still to great
extent in the ``taxonomic'' stage of development. We do not have any clear
classification and probably we have only discovered a small sample of
integrable difference equations, the tip of the iceberg.

As with differential equations, there is no universal definition of
integrability for discrete systems, and consensus about integrability
can exists only within certain subclasses of equations. In this
situation ``integrability predictors'' are very useful.  These are
algorithmic methods based on a somewhat weaker definition of
integrability, which nevertheless seem to be associated with
integrable systems. For example, for a Hamiltonian systems of $2N$
degrees of freedom ``Liouville integrability'' means the existence of
$N$ independent and sufficiently regular functions that commute
w.r.t.~the Poisson bracket. Proof of integrability would then require
the construction of the said commuting quantities, which is not
algorithmic. In this case a good algorithmic integrability predictor is
the Painlev\'e test.

\subsection{Integrability test and concepts in 1D}
For difference equations there is also a need for integrability
predictors. One such method is ``singularity confinement'' (SC)
\cite{Sing1}, which has been advocated as the discrete analogue of the
Painlev\'e test. The idea of this test is to check what happens at a
possible singularity of the evolution. Something special can only
happen, if the value of the dependent variable becomes infinite. A
singularity is then defined as a point where the next step cannot be
determined (for example due to expressions like $\infty - \infty$).
One then studies what happens near this point and the test condition
is the following: If the dynamics leads to (or near) a singularity
then after a few steps one should be able to get out of it, and this
should take place {\it without essential loss of information.}

This principle has been used successfully in deriving new integrable
difference equations, especially in finding discrete analogies of
Painlev\'e equations in \cite{Sing2} and in numerous subsequent papers
by many authors, especially trough the so-called ``de-autonomizing''
procedure. Despite its success, it turns out that passing the
singularity confinement test is {not} sufficient for regularity,
counterexamples are given in \cite{HV}.

It turns out that singularity confinement is strongly associated with
{\em reduced growth of complexity}. When one iterates a rational map
the expression becomes more and more complex as a rational expression
of the initial values. If we measure this complexity by the degree of
the numerator or denominator then generically the degree grows
exponentially. However, the growth can be reduced if some common
factors can be canceled.  Some amount of cancellation always happens
when singularity is confined\cite{HV2}, this is why SC is a such an
efficient test. However, as far as integrability is concerned, it is
the precise amount of cancellation that is crucial. This kind of
complexity analysis, alias algebraic entropy calculation has turned
out to be an unmatched integrability test for maps\cite{BeVi99,HV}.
The conjecture about growth after cancellations and integrability is
as follows:
\begin{itemize}
\item growth is linear in $n$  $\Rightarrow$ equation is linearizable.  
\item growth is polynomial in $n$  $\Rightarrow$ equation is
  integrable.
\item growth is exponential in $n$  $\Rightarrow$ equation is chaotic.  
\end{itemize}

\subsection{Integrability tests and concepts in 2D}
In principle the idea of singularity confinement can be applied to
lattice equations as well, this was already discussed in paper
\cite{Sing1} where the test was introduced. More recently this idea
was applied in \cite{SahaC} where an ``ultra-local singularity
confinement'' was proposed (see also \cite{SRH}). Nevertheless, it
turns out that SC is rarely used in the study of 2D maps, perhaps due
to the possibility of many different singularity confinement patterns
depending on arrangement of initial values. In contrast the growth of
complexity analysis can be easily applied to lattice equations as well
(see~\cite{TrGrRa01,Vi06}), and we will use it here.

Perhaps the strongest form of integrability is ``Consistency Around a
Cube'' as this kind of consistency immediately produces a Lax
pair~\cite{FLax,BSLax}.  The idea here is that one should be able to
consistently extend a two-dimensional map into three dimensions.  If
the 2D map is defined on an elementary square of the 2D lattice, then
one constructs a cube with a suitably modified maps on all sides of
the cube (the modification deals with associating different parameters
to different coordinate directions).  This creates a potential
consistency problem in the evolution as follows: suppose we are given
the values at the corners $x_{000},\,x_{100},\,x_{010},\,x_{001}$
(initial values), then the values at $x_{110},\,x_{101},\,x_{011}$ are
uniquely determined by using the proper maps, but the value at
$x_{111}$ can be computed in 3 different ways. The consistency test is
that these three ways should all yield the same value.  This is a kind
of Bianchi identity.  It has been used as a method to find and classify
integrable lattice models, when associated with rather strong symmetry
requirements and with \cite{ABS} or without \cite{JHside} the so
called ``tetrahedron property''.

However, this kind of consistency is very sensitive on the role of
{\em spectral parameters}: the coefficients of the model have specific
roles and are interdependent. This means, for example, that the normal
set of linear transformations would completely destroy any such
association.  Unfortunately imposing this sort of interdependency adds
enormous complications for a generic search program.

\subsection{Plan of the paper}
We start from lattice relations defined on an elementary square of the
2D lattice. We assume the relation is at most quadratic in the
dependent variables.  Due to the reasons discussed above we have
chosen factorization as the first selecting criterion: we impose as a
factorization requirement that (at least) a linear factor can be
extracted after 2 steps of evolution. Solving the resulting equations
produces a list of maps, as described in Section \ref{S2}.

There are natural symmetries between the models. For example two
models may be related by a translation of the variables. Maps related
by reflections or rotations can also be omitted, and this allows to
reduce the list to 80 cases.

We then analyze all 80 cases with an algebraic entropy calculation, as
explained is Section \ref{S3}. This yields a classification of the
models according to their degree growth. The vanishing of entropy
(polynomial growth) is the integrability detector we use here.
However, since our search condition is just the factorization of one
linear factor after two iterations, many models with exponential
growth are still included, which may contain integrable models, when
further constraints are introduced. Our final list contains just the
different ``parents'' without further analysis on their possible
integrable ``descendents''.

In section \ref{S:skkn}, we analyze in with some detail one
particularly interesting multiparametric model, and show that it also
verifies the condition of consistency around the cube, thus proving
its integrability. This model does not have the symmetries used
in~\cite{ABS}, and thus does not appear in the list given there, but as
special cases it contains both the discrete modified and Schwarzian KdV.

\section{Factorization in 2D}\label{S1}
\subsection{The map}
\begin{figure}
\setlength{\unitlength}{0.00035in}
\begin{picture}(3482,3500)(500,-200)

\put(3275,2708){\circle*{150}}
\put(2625,2808){\makebox(0,0)[lb]{$x_{[2]}$}}

\put(5075,2708){\circle*{150}}
\put(5375,2808){\makebox(0,0)[lb]{$x_{[12]}$}}

\put(3275,908){\circle*{150}}
\put(2825,1008){\makebox(0,0)[lb]{$x$}}

\put(5075,908){\circle*{150}}
\put(5375,1008){\makebox(0,0)[lb]{$x_{[1]}$}}

\drawline(2275,2708)(6075,2708)
\drawline(5075,3633)(5075,0)
\drawline(2275,908)(6075,908)
\drawline(3275,3633)(3275,0)
\end{picture}
\caption{The map is defined on an elementary square of the 2D
  lattice.}
\label{F1}
\end{figure}
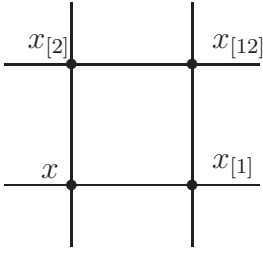

We consider maps defined on the cartesian 2D lattice by relating the
four corner values of an elementary square (see Figure \ref{F1}) with
a multilinear relation: 
\begin{align}
  & k\, x x_{[1]} x_{[2]} x_{[12]} + { l_1\, x x_{[1]} x_{[2]} + l_2\,
    x x_{[1]} x_{[12]}
    + l_3\, x x_{[2]} x_{[12]} + l_4\, x_{[1]} x_{[2]} x_{[12]}}\nonumber\\
  & + { p_1\, x x_{[1]} + p_2\, x_{[1]} x_{[2]} + p_3\, x_{[2]}
    x_{[12]} + p_4\, x_{[12]} x +
    p_5\, x x_{[2]} + p_6\, x_{[1]} x_{[12]}}\nonumber\\
  & + { r_1\, x + r_2 \,x_{[1]} + r_3\, x_{[2]} + r_4\, x_{[12]}} +
  u\equiv {Q(x,x_{[1]},x_{[2]},x_{[12]};\alpha_1,\alpha_2)\!=\!0}.\label{M}
\end{align}
Here $x_{n,m}$ is the dependent variable at a corner and we have used
a shorthand notation, in which only the shifts with respect to the
base point at lower left is indicated in a subscript in square
brackets: $x_{n,m}=x_{00}=x$, $x_{n+1,m}=x_{10}=x_{[1]}$,
$x_{n,m+1}=x_{01}=x_{[2]}$, $x_{n+1,m+1}=x_{11}=x_{[12]}$. (We
  will use indifferently these three notations).  As
mentioned above, in the 3D consistency approach an important role is
played by the spectral parameters $\alpha_s$, associated to specific
directions, and they appear in the coefficients $k,\,l_i,\,
p_i,\,r_i,\,u$. In this paper, however, these spectral parameters are
ignored (except in Section \ref{S:skkn}), since we only work on the 2D
lattice.

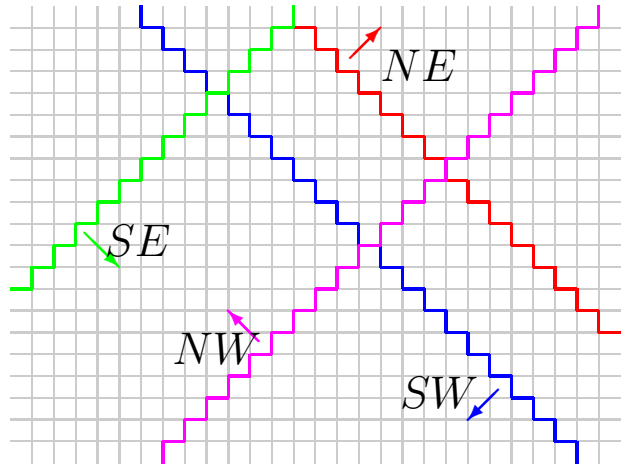
\begin{figure}[b]
\begin{center}
\setlength{\unitlength}{1200sp}%
\begingroup\makeatletter\ifx\SetFigFont\undefined%
\gdef\SetFigFont#1#2#3#4#5{%
  \reset@font\fontsize{#1}{#2pt}%
  \fontfamily{#3}\fontseries{#4}\fontshape{#5}%
  \selectfont}%
\fi\endgroup%
\begin{picture}(12708,8000)(397,-8000)
\label{fourcorners}
\thinlines
{\color[rgb]{.8,.8,.8}\put(901,389){\line( 0,-1){9450}}
}%
{\color[rgb]{.8,.8,.8}\put(1351,389){\line( 0,-1){9450}}
}%
{\color[rgb]{.8,.8,.8}\put(1801,389){\line( 0,-1){9450}}
}%
{\color[rgb]{.8,.8,.8}\put(2251,389){\line( 0,-1){9450}}
}%
{\color[rgb]{.8,.8,.8}\put(2701,389){\line( 0,-1){9450}}
}%
{\color[rgb]{.8,.8,.8}\put(3151,389){\line( 0,-1){9450}}
}%
{\color[rgb]{.8,.8,.8}\put(3601,389){\line( 0,-1){9450}}
}%
{\color[rgb]{.8,.8,.8}\put(4051,389){\line( 0,-1){9450}}
}%
{\color[rgb]{.8,.8,.8}\put(4501,389){\line( 0,-1){9450}}
}%
{\color[rgb]{.8,.8,.8}\put(4951,389){\line( 0,-1){9450}}
}%
{\color[rgb]{.8,.8,.8}\put(5401,389){\line( 0,-1){9450}}
}%
{\color[rgb]{.8,.8,.8}\put(5851,389){\line( 0,-1){9450}}
}%
{\color[rgb]{.8,.8,.8}\put(6301,389){\line( 0,-1){9450}}
}%
{\color[rgb]{.8,.8,.8}\put(6751,389){\line( 0,-1){9450}}
}%
{\color[rgb]{.8,.8,.8}\put(7201,389){\line( 0,-1){9450}}
}%
{\color[rgb]{.8,.8,.8}\put(7651,389){\line( 0,-1){9450}}
}%
{\color[rgb]{.8,.8,.8}\put(8101,389){\line( 0,-1){9450}}
}%
{\color[rgb]{.8,.8,.8}\put(8551,389){\line( 0,-1){9450}}
}%
{\color[rgb]{.8,.8,.8}\put(9001,389){\line( 0,-1){9450}}
}%
{\color[rgb]{.8,.8,.8}\put(9451,389){\line( 0,-1){9450}}
}%
{\color[rgb]{.8,.8,.8}\put(9901,389){\line( 0,-1){9450}}
}%
{\color[rgb]{.8,.8,.8}\put(10351,389){\line( 0,-1){9450}}
}%
{\color[rgb]{.8,.8,.8}\put(10801,389){\line( 0,-1){9450}}
}%
{\color[rgb]{.8,.8,.8}\put(11251,389){\line( 0,-1){9450}}
}%
{\color[rgb]{.8,.8,.8}\put(11701,389){\line( 0,-1){9450}}
}%
{\color[rgb]{.8,.8,.8}\put(12151,389){\line( 0,-1){9450}}
}%
{\color[rgb]{.8,.8,.8}\put(12601,389){\line( 0,-1){9450}}
}%
{\color[rgb]{.8,.8,.8}\put(451,-61){\line( 1, 0){12600}}
}%
{\color[rgb]{.8,.8,.8}\put(451,-61){\line( 1, 0){12600}}
}%
{\color[rgb]{.8,.8,.8}\put(451,-61){\line( 1, 0){12600}}
}%
{\color[rgb]{.8,.8,.8}\put(451,-61){\line( 1, 0){12600}}
}%
{\color[rgb]{.8,.8,.8}\put(451,-511){\line( 1, 0){12600}}
}%
{\color[rgb]{.8,.8,.8}\put(451,-961){\line( 1, 0){12600}}
}%
{\color[rgb]{.8,.8,.8}\put(451,-1411){\line( 1, 0){12600}}
}%
{\color[rgb]{.8,.8,.8}\put(451,-1861){\line( 1, 0){12600}}
}%
{\color[rgb]{.8,.8,.8}\put(451,-2311){\line( 1, 0){12600}}
}%
{\color[rgb]{.8,.8,.8}\put(451,-2761){\line( 1, 0){12600}}
}%
{\color[rgb]{.8,.8,.8}\put(451,-3211){\line( 1, 0){12600}}
}%
{\color[rgb]{.8,.8,.8}\put(451,-3661){\line( 1, 0){12600}}
}%
{\color[rgb]{.8,.8,.8}\put(451,-4111){\line( 1, 0){12600}}
}%
{\color[rgb]{.8,.8,.8}\put(451,-4561){\line( 1, 0){12600}}
}%
{\color[rgb]{.8,.8,.8}\put(451,-5011){\line( 1, 0){12600}}
}%
{\color[rgb]{.8,.8,.8}\put(451,-5461){\line( 1, 0){12600}}
}%
{\color[rgb]{.8,.8,.8}\put(451,-5911){\line( 1, 0){12600}}
}%
{\color[rgb]{.8,.8,.8}\put(451,-6361){\line( 1, 0){12600}}
}%
{\color[rgb]{.8,.8,.8}\put(451,-6811){\line( 1, 0){12600}}
}%
{\color[rgb]{.8,.8,.8}\put(451,-7261){\line( 1, 0){12600}}
}%
{\color[rgb]{.8,.8,.8}\put(451,-7711){\line( 1, 0){12600}}
}%
{\color[rgb]{.8,.8,.8}\put(451,-8161){\line( 1, 0){12600}}
}%
{\color[rgb]{.8,.8,.8}\put(451,-8611){\line( 1, 0){12600}}
}%
\thicklines
{\color[rgb]{1,0,0}\put(6301,389){\line( 0,-1){450}}
\put(6301,-61){\line( 1, 0){450}}
\put(6751,-61){\line( 0,-1){450}}
\put(6751,-511){\line( 1, 0){450}}
\put(7201,-511){\line( 0,-1){450}}
\put(7201,-961){\line( 1, 0){450}}
\put(7651,-961){\line( 0,-1){450}}
\put(7651,-1411){\line( 1, 0){450}}
\put(8101,-1411){\line( 0,-1){450}}
\put(8101,-1861){\line( 1, 0){450}}
\put(8551,-1861){\line( 0,-1){450}}
\put(8551,-2311){\line( 1, 0){450}}
\put(9001,-2311){\line( 0,-1){450}}
\put(9001,-2761){\line( 1, 0){450}}
\put(9451,-2761){\line( 0,-1){450}}
\put(9451,-3211){\line( 1, 0){450}}
\put(9901,-3211){\line( 0,-1){450}}
\put(9901,-3661){\line( 1, 0){450}}
\put(10351,-3661){\line( 0,-1){450}}
\put(10351,-4111){\line( 1, 0){450}}
\put(10801,-4111){\line( 0,-1){450}}
\put(10801,-4561){\line( 1, 0){450}}
\put(11251,-4561){\line( 0,-1){450}}
\put(11251,-5011){\line( 1, 0){450}}
\put(11701,-5011){\line( 0,-1){450}}
\put(11701,-5461){\line( 1, 0){450}}
\put(12151,-5461){\line( 0,-1){450}}
\put(12151,-5911){\line( 1, 0){450}}
\put(12601,-5911){\line( 0,-1){450}}
\put(12601,-6361){\line( 1, 0){450}}
}%
{\color[rgb]{0,0,1}\put(3151,389){\line( 0,-1){450}}
\put(3151,-61){\line( 1, 0){450}}
\put(3601,-61){\line( 0,-1){450}}
\put(3601,-511){\line( 1, 0){450}}
\put(4051,-511){\line( 0,-1){450}}
\put(4051,-961){\line( 1, 0){450}}
\put(4501,-961){\line( 0,-1){450}}
\put(4501,-1411){\line( 1, 0){450}}
\put(4951,-1411){\line( 0,-1){450}}
\put(4951,-1861){\line( 1, 0){450}}
\put(5401,-1861){\line( 0,-1){450}}
\put(5401,-2311){\line( 1, 0){450}}
\put(5851,-2311){\line( 0,-1){450}}
\put(5851,-2761){\line( 1, 0){450}}
\put(6301,-2761){\line( 0,-1){450}}
\put(6301,-3211){\line( 1, 0){450}}
\put(6751,-3211){\line( 0,-1){450}}
\put(6751,-3661){\line( 1, 0){450}}
\put(7201,-3661){\line( 0,-1){450}}
\put(7201,-4111){\line( 1, 0){450}}
\put(7651,-4111){\line( 0,-1){450}}
\put(7651,-4561){\line( 1, 0){450}}
\put(8101,-4561){\line( 0,-1){450}}
\put(8101,-5011){\line( 1, 0){450}}
\put(8551,-5011){\line( 0,-1){450}}
\put(8551,-5461){\line( 1, 0){450}}
\put(9001,-5461){\line( 0,-1){450}}
\put(9001,-5911){\line( 1, 0){450}}
\put(9451,-5911){\line( 0,-1){450}}
\put(9451,-6361){\line( 1, 0){450}}
\put(9901,-6361){\line( 0,-1){450}}
\put(9901,-6811){\line( 1, 0){450}}
\put(10351,-6811){\line( 0,-1){450}}
\put(10351,-7261){\line( 1, 0){450}}
\put(10801,-7261){\line( 0,-1){450}}
\put(10801,-7711){\line( 1, 0){450}}
\put(11251,-7711){\line( 0,-1){450}}
\put(11251,-8161){\line( 1, 0){450}}
\put(11701,-8161){\line( 0,-1){450}}
\put(11701,-8611){\line( 1, 0){450}}
\put(12151,-8611){\line(-1, 0){450}}
\put(11701,-8611){\line( 1, 0){450}}
\put(12151,-8611){\line( 0,-1){450}}
}%
{\color[rgb]{0,1,0}\put(451,-5461){\line( 1, 0){450}}
\put(901,-5461){\line( 0, 1){450}}
\put(901,-5011){\line( 1, 0){450}}
\put(1351,-5011){\line( 0, 1){450}}
\put(1351,-4561){\line( 1, 0){450}}
\put(1801,-4561){\line( 0, 1){450}}
\put(1801,-4111){\line( 1, 0){450}}
\put(2251,-4111){\line( 0, 1){450}}
\put(2251,-3661){\line( 1, 0){450}}
\put(2701,-3661){\line( 0, 1){450}}
\put(2701,-3211){\line( 1, 0){450}}
\put(3151,-3211){\line( 0, 1){450}}
\put(3151,-2761){\line( 1, 0){450}}
\put(3601,-2761){\line( 0, 1){450}}
\put(3601,-2311){\line( 1, 0){450}}
\put(4051,-2311){\line( 0, 1){450}}
\put(4051,-1861){\line( 1, 0){450}}
\put(4501,-1861){\line( 0, 1){450}}
\put(4501,-1411){\line( 1, 0){450}}
\put(4951,-1411){\line( 0, 1){450}}
\put(4951,-961){\line( 1, 0){450}}
\put(5401,-961){\line( 0, 1){450}}
\put(5401,-511){\line( 1, 0){450}}
\put(5851,-511){\line( 0, 1){450}}
\put(5851,-61){\line( 1, 0){450}}
\put(6301,-61){\line( 0, 1){450}}
}%
{\color[rgb]{1,0,1}\put(12601,389){\line( 0,-1){450}}
\put(12601,-61){\line(-1, 0){450}}
\put(12151,-61){\line( 0,-1){450}}
\put(12151,-511){\line(-1, 0){450}}
\put(11701,-511){\line( 0,-1){450}}
\put(11701,-961){\line(-1, 0){450}}
\put(11251,-961){\line( 0,-1){450}}
\put(11251,-1411){\line(-1, 0){450}}
\put(10801,-1411){\line( 0,-1){450}}
\put(10801,-1861){\line(-1, 0){450}}
\put(10351,-1861){\line( 0,-1){450}}
\put(10351,-2311){\line(-1, 0){450}}
\put(9901,-2311){\line( 0,-1){450}}
\put(9901,-2761){\line(-1, 0){450}}
\put(9451,-2761){\line( 0,-1){450}}
\put(9451,-3211){\line(-1, 0){450}}
\put(9001,-3211){\line( 0,-1){450}}
\put(9001,-3661){\line(-1, 0){450}}
\put(8551,-3661){\line( 0,-1){450}}
\put(8551,-4111){\line(-1, 0){450}}
\put(8101,-4111){\line( 0,-1){450}}
\put(8101,-4561){\line(-1, 0){450}}
\put(7651,-4561){\line( 0,-1){450}}
\put(7651,-5011){\line(-1, 0){450}}
\put(7201,-5011){\line( 0,-1){450}}
\put(7201,-5461){\line(-1, 0){450}}
\put(6751,-5461){\line( 0,-1){450}}
\put(6751,-5911){\line(-1, 0){450}}
\put(6301,-5911){\line( 0,-1){450}}
\put(6301,-6361){\line(-1, 0){450}}
\put(5851,-6361){\line( 0,-1){450}}
\put(5851,-6811){\line(-1, 0){450}}
\put(5401,-6811){\line( 0,-1){450}}
\put(5401,-7261){\line(-1, 0){450}}
\put(4951,-7261){\line( 0,-1){450}}
\put(4951,-7711){\line(-1, 0){450}}
\put(4501,-7711){\line( 0,-1){450}}
\put(4501,-8161){\line(-1, 0){450}}
\put(4051,-8161){\line( 0,-1){450}}
\put(4051,-8611){\line(-1, 0){450}}
\put(3601,-8611){\line( 0,-1){450}}
}%
{\color[rgb]{1,0,1}\put(5581,-6541){\line(-1, 1){630}}
}%
{\color[rgb]{1,0,1}\put(5581,-6541){\vector(-1, 1){630}}
}%
{\color[rgb]{0,1,0}\put(1981,-4291){\vector( 1,-1){720}}
}%
{\color[rgb]{1,0,0}\put(7471,-691){\vector( 1, 1){630}}
}%
{\color[rgb]{0,0,1}\put(10531,-7531){\vector(-1,-1){630}}
}%
\put(3800,-6811){\makebox(0,0)[lb]{\smash{{\SetFigFont{25}{30.0}{\rmdefault}{\mddefault}{\updefault}{\color[rgb]{0,0,0}${}_{NW}$}%
}}}}
\put(2400,-4561){\makebox(0,0)[lb]{\smash{{\SetFigFont{25}{30.0}{\rmdefault}{\mddefault}{\updefault}{\color[rgb]{0,0,0}${}_{SE}$}%
}}}}
\put(8101,-961){\makebox(0,0)[lb]{\smash{{\SetFigFont{25}{30.0}{\rmdefault}{\mddefault}{\updefault}{\color[rgb]{0,0,0}${}_{NE}$}%
}}}}
\put(8500,-7711){\makebox(0,0)[lb]{\smash{{\SetFigFont{25}{30.0}{\rmdefault}{\mddefault}{\updefault}{\color[rgb]{0,0,0}${}_{SW}$}%
}}}}
\end{picture}%
\end{center}
\caption{Fundamental evolutions on a square lattice.}
\label{F:comp}
\end{figure}

Dynamics is defined by $Q(x_{n,m},x_{n+1,m},x_{n,m+1},x_{n+1,m+1})=0$,
using the multilinear $Q$ of \eqref{M}, and this allows well-defined
evolution from any staircase-like initial condition, up or down, since
we can solve for any particular corner value in terms of the others,
see Figure \ref{F:comp}.

The canonical examples of such maps are the following:
\begin{itemize}
\item Lattice KdV
\[
(p_1-p_2+x_{0 1}-x_{1 0})(p_1+p_2+x_{0 0}-x_{1 1})=p_1^2-p_2^2,
\]
or after translation $x_{n,m}=u_{n,m}+p_1n+p_2m$
\begin{equation}\label{kdv}
(u_{0 1}-u_{1 0})(u_{0 0}-u_{1 1})=p_1^2-p_2^2,
\end{equation}

\item Lattice MKdV
\begin{equation}\label{mkdv}
p_1(x_{0 0} x_{0 1}-x_{1 0}x_{1 1})=p_2(x_{0 0}
x_{1 0}-x_{0 1}x_{1 1}),
\end{equation}

\item Lattice SKdV
\begin{equation}\label{skdv}
(x_{0 0}-x_{1 0})(x_{0 1}-x_{1 1})p_1^2=
(x_{0 0}-x_{0 1})(x_{1 0}-x_{1 1})p_2^2.
\end{equation}

\end{itemize}

\subsection{Keeping track of factors}
When a rational map is iterated some factors often get canceled
``silently'' in the resulting rational expression. This can be
observed by comparing the degrees of the terms to the generic case.
However, the best way to keep track of factors it to write the
rational map as a polynomial map in a projective space.  This was also
the method that in the 1D-case showed clearly the connection between
singularity confinement and reduction in the growth of
complexity\cite{HV2}.  In the 1D-case it turned out that the amount of
cancellation had to be such that the degrees only grow polynomially
($\propto \; n^k$), while for nonintegrable systems we have exponential
growth ($\rho^n$ with $\rho > 1$). The same happens for
2D-systems\cite{TrGrRa01,Vi06}.

Recall that the map we are considering is given by a multilinear expression
\[
Q(x_{n,m},x_{n+1,m},x_{n,m+1},x_{n+1,m+1})=0.
\]
This equation can be homogenized by substituting
$x_{n,m}=v_{n,m}/f_{n,m}$ and taking the numerator. We assume that $Q$
does not factor and that it depends on all the indicated variables. In
the numerator of the homogenized $Q$ we isolate $v_{n+1,m+1}$ and
$f_{n+1,m+1}$:
\[
A({\scriptstyle
  v_{n,m},v_{n+1,m},v_{n,m+1},f_{n,m},f_{n+1,m},f_{n,m+1}})\,
v_{n+1,m+1}
+B({\scriptstyle
  v_{n,m},v_{n+1,m},v_{n,m+1},f_{n,m},f_{n+1,m},f_{n,m+1}}) \,f_{n+1,m+1}
\]
and then define the projective map as
\begin{equation}
  \left\{\begin{array}{rcl} 
v_{n+1,m+1}&=&-B({\scriptstyle v_{n,m},v_{n+1,m},v_{n,m+1}, 
  f_{n,m},f_{n+1,m},f_{n,m+1}}),\\
f_{n+1,m+1}&=&A({\scriptstyle v_{n,m},v_{n+1,m},v_{n,m+1},
  f_{n,m},f_{n+1,m},f_{n,m+1}}).
\end{array}\right.
\label{projmap}
\end{equation}
The polynomials $A,B$ are both of degree 3 in the indicated variables,
and they cannot have common factors, since $Q$ was assumed
irreducible.

For practical reasons we consider in this paper {\em only quadratic}
maps, i.e., in \eqref{M} we take $k=0,\, l_i=0$. In that case the
projective map is given by
\begin{equation}
\left\{\begin{array}{rcl} 
v_{[12]}&=&-[p_1\, v\, v_{[1]}f_{[2]} + p_2\,
    v_{[1]} v_{[2]}f
    + p_5\, v\,  v_{[2]}f_{[1]}\\
    && \quad + r_1\, v\, f_{[1]}f_{[2]} + r_2 \,v_{[1]}f_{[2]}f +
    r_3\, v_{[2]}f_{[1]}f + u\,  f\,  f_{[1]}f_{[2]}],\\
    f_{[12]}&=& p_3\, v_{[2]}f_{[1]}f + p_4\, v\, f_{[1]}f_{[2]} +
    p_6\, v_{[1]}f_{[2]}f + r_4f\, f_{[1]}f_{[2]}.
\end{array}\right.
\label{promap}
\end{equation}
From this we see immediately that the default degree growth is:
\begin{equation}\label{gro}
\deg(z_{n+1,m+1})=\deg(z_{n+1,m})+\deg(z_{n,m+1})+\deg(z_{n,m}),
\end{equation}
where $z=v$ or $f$, since they have the same degree.

For the initial values we may take any staircase like configuration
and in the rational representation take arbitrary $x$'s at each point.
In the projective representation we may also take arbitrary $v$'s at
all initial points, but we should use the same $f$, because
projectivity only adds one free overall factor. 
In the quadratic case this means the cancellation of one extra term at
each step, so that the sequence of maximal degrees (without any
factorization) is $ 1, 2, 4, 9, 21, \dots$ which corresponds to the
asymptotically exponential growth $(1+\sqrt{2})^k$.

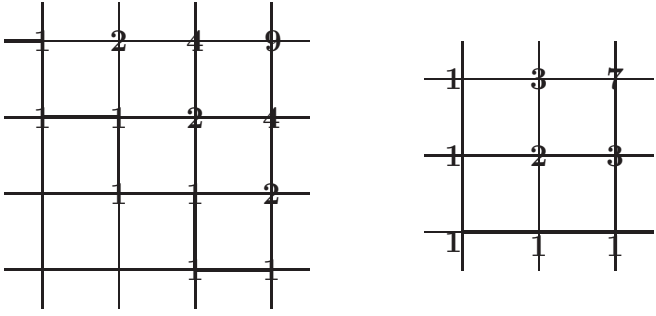
\begin{figure}
\setlength{\unitlength}{0.004in}
\begin{picture}(300,350)(-100, 200)
\drawline(100,200)(500,200)
\drawline(100,300)(500,300)
\drawline(100,400)(500,400)
\drawline(100,500)(500,500)
\drawline(150,550)(150,150)
\drawline(250,550)(250,150)
\drawline(350,550)(350,150)
\drawline(450,550)(450,150)
\put(150,500){\makebox(0,0)[cc]{\bf 1}}
\put(150,400){\makebox(0,0)[cc]{\bf 1}}
\put(250,400){\makebox(0,0)[cc]{\bf 1}}
\put(250,300){\makebox(0,0)[cc]{\bf 1}}
\put(350,300){\makebox(0,0)[cc]{\bf 1}}
\put(350,200){\makebox(0,0)[cc]{\bf 1}}
\put(450,200){\makebox(0,0)[cc]{\bf 1}}
\put(250,500){\makebox(0,0)[cc]{\bf 2}}
\put(350,400){\makebox(0,0)[cc]{\bf 2}}
\put(450,300){\makebox(0,0)[cc]{\bf 2}}
\put(350,500){\makebox(0,0)[cc]{\bf 4}}
\put(450,400){\makebox(0,0)[cc]{\bf 4}}
\put(460,500){\makebox(0,0)[cc]{{\bf 9 }}}
\thicklines
\drawline(100,500)(150,500)
\drawline(150,500)(150,400)
\drawline(150,400)(250,400)
\drawline(250,400)(250,300)
\drawline(250,300)(350,300)
\drawline(350,300)(350,200)
\drawline(350,200)(450,200)
\drawline(450,200)(450,150)
\end{picture}

\begin{picture}(600,0)(-550,200)
\drawline(200,300)(500,300)
\drawline(200,400)(500,400)
\drawline(200,500)(500,500)
\drawline(250,550)(250,250)
\drawline(350,550)(350,250)
\drawline(450,550)(450,250)
\put(250,400){\makebox(0,0)[cr]{\bf 1}}
\put(250,300){\makebox(0,0)[tr]{\bf 1}}
\put(350,295){\makebox(0,0)[tc]{\bf 1}}
\put(250,500){\makebox(0,0)[rc]{\bf 1}}
\put(450,295){\makebox(0,0)[tc]{\bf 1}}
\put(350,400){\makebox(0,0)[cc]{\bf 2}}
\put(350,500){\makebox(0,0)[cc]{\bf 3}}
\put(450,400){\makebox(0,0)[cc]{\bf 3}}
\put(450,500){\makebox(0,0)[cc]{{\bf 7}}}
\thicklines

\drawline(250,550)(250,300)
\drawline(250,300)(500,300)
\end{picture}

\caption{Maximum degrees with staircase and corner initial states for
  quadratic relations.\label{F2}}
\end{figure}

For the search part of our factorization study we will only consider
the first few steps in the iteration of \eqref{promap}. In principle
two different initial configurations are often used, staircase and
corner, the default degrees in these two cases are given in
Fig.~\ref{F2}. The interesting factorization, that can be used to
predict integrability, happens at the point where the default degree
shown in Fig.~\ref{F2} is 9 or 7, respectively.  In the search part we
use the corner configuration, because the total degree in then
smallest. In the degree growth analysis we have used the staircase
configuration.

\subsection{Factorization of known models}
\label{known_models}
Consider the discrete KdV model given by
\[
(x_{n, m+1}-x_{n+1,m})(x_{n,m }-x_{n+1,m+ 1})=a,
\]
The homogenized $Q$ is now
\begin{align*}
Q=&f_{n,m+1} f_{n,m} v_{n+1,m+1} v_{n+1,m} - f_{n+1,m} f_{n,m} 
v_{n+1,m+1} v _{n,m+1} - f_{n+1,m+1} f_{n,m+1} v_{n+1,m} v_{n,m}\\
 &+ f_{n+1,m+1} f_{n+1,m} v_{n,m+1} v_{n,m} - f_{n+1,m+1} f_ {n+1,m}
f_{n,m+1} f_{n,m} a,
\end{align*}
and the projective map
\begin{equation}
  \left\{\begin{array}{rcl} 
      v_{n+1,m+1}&=&f_{n,m+1} v_{n+1,m} v_{n,m} - 
      f_{n+1,m} v_{n,m+1} v_{n,m} + f_{n+1,m} f_{n,m+1} f_{n,m} a,\\
      f_{n+1,m+1}&=&f_{n,m+1} f_{n,m} v_{n+1,m} - 
      f_{n+1,m} f_{n,m} v_{n,m+1}.
\end{array}\right.
\label{kdvprojmap}
\end{equation}
In the corner case we take $f_{0,m}=f_{n,0}=f_{00}$ and after
canceling one $f_{00}$ at every step one find the first interesting GCD
(greatest common divisor of the list of polynomials) at
$(2,2)$:
\[
{\rm GCD}(v_{22},f_{22})=(v_{01}-v_{10})^2.
\]
Exactly the same GCD is found from the stair configuration, with
initial values restricted by $f_{n,m}=f_{00}$ for $n+m=0,1$. The remaining
(different) factors of $v_{22},f_{22}$ are rather lengthy, in the
corner case they are of degree 5 in the initial values and have 23 and
13 terms, respectively, and for the staircase initial configuration
the degree is 7 and the number of terms 112 and 76, respectively.

The simplest linear model $x_{[12]}+x_{[1]}+x_{[2]}+x=0$ the
cancellations are so strong that the map stay linear, similarly for
the quadratic model $x_{[12]}x+ax_{[1]}x_{[2]}=0$ the projective map
stays quadratic.

In \cite{ABS} a list of lattice models having the ``consistency around a cube'' property was
given. The divisors for the quadratic cases, using corner
configuration, are as follows

\begin{itemize}
\item[(H1)]
map: $(x_{00}-x_{11}) (x_{10}-x_{01})+\beta-\alpha =0,$

factor $(v_{10} - v_{01})^2.$

\item[(H2)] map: $(x_{00}-x_{11}) (x_{10}-x_{01})+
(\beta-\alpha ) (x_{00}+x_{10}+x_{01}+x_{11})+\beta^2-\alpha ^2 =0,$

factor: $(v_{10} - v_{01} + (\alpha - \beta)f_{00})
(v_{10} - v_{01} - (\alpha - \beta)f_{00})$.

\item[(H3)]
map: $\alpha  (x_{00} x_{10}+x_{01} x_{11})-
\beta (x_{00} x_{01}+x_{10} x_{11})+\delta (\alpha ^2-\beta^2) =0$,

factor: $ (v_{10} \alpha- v_{01} \beta)(v_{01} \alpha - v_{10} \beta)$.

\item[(A1)]
map: $\alpha (x_{00}+x_{01}) (x_{10}+x_{11}) - \beta(x_{00}+x_{10})
(x_{01}+x_{11})- \delta^2 \alpha  \beta (\alpha -\beta) =0$,

factor:  $(v_{10} - v_{01} + \delta(\alpha - \beta)f_{00})
(v_{10} - v_{01} - \delta(\alpha - \beta)f_{00})$.

\item[(Q1)]
map: $\alpha  (x_{00}-x_{01}) (x_{10}-x_{11})-
\beta (x_{00}-x_{10}) (x_{01}-x_{11})+\delta^2 \alpha  \beta (\alpha
-\beta) =0$, 

factor: $(v_{10} - v_{01} + \delta(\alpha - \beta)f_{00})
(v_{10} - v_{01} - \delta(\alpha - \beta)f_{00})$.

\item[(Q2)] map: $\alpha  (x_{00}-x_{01}) (x_{10}-x_{11})-
\beta (x_{00}-x_{10}) (x_{01}-x_{11})\\ +
\alpha  \beta (\alpha -\beta) (x_{00}+x_{10}+x_{01}+x_{11})
-\alpha  \beta (\alpha -\beta) (\alpha ^2-\alpha  \beta+\beta^2) =0$,

factor: $(v_{10}-v_{01})^2-2 (\alpha-\beta)^2 f_{00}
(v_{10}+v_{01})+(\alpha-\beta)^4 f_{00}^2$.

\item[(Q3)] map: $(\beta^2-\alpha ^2) (x_{00} x_{11}+x_{10} x_{01})
+\beta (\alpha ^2-1) (x_{00} x_{10}+x_{01} x_{11})\\
-\alpha  (\beta^2-1) (x_{00} x_{01}+x_{10} x_{11})
-\delta^2 (\alpha ^2-\beta^2) (\alpha ^2-1) (\beta^2-1)/(4 \alpha  \beta) =0$,

factor: $ 4 \alpha \beta (v_{01} \alpha - v_{10} \beta) (v_{01} \beta
- v_{10} \alpha)+ \delta^2 (\alpha^2-\beta^2)^2f_{00}^2$.

\end{itemize}
Note that for H1-H3, A1, Q1 we get two linear factors while for Q2, Q3
the quadratic factor is irreducible (if $\delta=0$ then the divisor
for Q3 does factor). Note also that the common factors can be used to
study relationships between maps; for the above list we just observe that
A1 is obtained from Q1 with $x_{nm}\mapsto (-1)^{n+m}x_{nm}$.

\section{Search}\label{S2}
\subsection{The method}
We have seen above that in all known integrable quadratic lattice maps
the result at $(2,2)$ factorizes with a quadratic GCD (with the
exception of linearizable models, that may have even more factors). In
many models this quadratic common divisor factorizes, but not always.
The search for a linear factor is computationally easier, and
therefore this search is restricted to that, but hope to return to a
search for model with irreducible quadratic factors later.  The
possibility of irreducible factors of degree higher than 2 is open,
and no examples are known.

In this search project we use the corner configuration, because
computations are then simpler. Also, for computational simplicity, we
use $x$ rather than $v,f$, but to prevent accidental factorizations we
proceed as follows: we calculate $x_{11},\, x_{21},\, x_{12}$ using a
{\em generic} $Q$ (which does not factorize), and substitute the obtained
values into the equation $Q(x_{11},x_{12},x_{21},x_{22})=0$ and take
its numerator (to be called $Q_N$ from now on). We then require that
this $Q_N$ factorizes as $Q_N=(\text{some polynomial in } x_{00},
x_{10}, x_{01},x_{20},x_{02},x_{22}) \times
\left[x_{01}+s_{10}x_{10}+s_{00}\right]$. Here we may assume that
$x_{01}$ has unit coefficient, because the other coefficients $s$ can
be rational in $x_{00}$, furthermore, if $x_{01}$ were zero but
$x_{10}$ not we could use $n\leftrightarrow m$ reflection.

If we keep track of all the other variables except $x_{00}$ we have
generically
\begin{equation}
Q_N=\sum_{\alpha ,\mu ,\nu =0}^1
\sum_{t,v=0}^4
x_{22}^{\alpha } x_{20}^{\mu } x_{02}^{\nu } 
x_{10}^{t} x_{01}^{v}g(x_{00},\alpha  ,\nu ,\mu  ,t
,v)
\end{equation}
for some polynomial $g$, furthermore we know something about the exponents:
$\alpha +\mu +\nu +t+v\le7$. Because of this observation we can make 
the following ansatz for the factorization $Q=PS$:
\begin{align}
  S=&x_{01}+s_{10}x_{10}+s_{00},\\
  P=&\sum_{\alpha ,\mu ,\nu =0}^1 \sum_{t,v=0}^4 x_{22}^{\alpha }
  x_{20}^{\mu } x_{02}^{\nu } x_{10}^{t} x_{01}^{v}d(x_{00},\alpha
  ,\nu ,\mu ,t ,v),
\end{align}
where $\alpha +\mu +\nu +t+v\le6$, and $d$, $s$ are rational in
$x_{00}$.

Given the ansatz \eqref{M} with the quadratic restriction
$k=0,\,l_i=0$ we compute the $Q_N$ (which has 15966 terms) and
subtract the ansatz $PS$ above.  The equations are then formed by
taking the coefficients of different powers of $x_{22}, x_{20},
x_{02}, x_{10}, x_{01}$ with fixed ordering. Next we determined the
134 functions $d$ in $P$ by considering the leading terms of
$Q-P(x_{01}+\dots)=0$. This is fully automatized using
REDUCE\cite{RED}.  The remaining equations were then solved for the
functions $s$ and for the parameters of the map itself, $p_i,r_i,u$.
The solution process branched a lot and required numerous separate
computing sessions.

\section{Results}\label{S3}
If the computations led to a situation where the resulting map a)
factorized, or b) did not depend on all corners, that branch was
terminated immediately, but even then we got 125 ``raw'' results. From
this set we omitted maps that could be obtained by rotation or
reflection from other maps (using translation by a constant if
needed), this reduced the number from 125 to 80. For all those
remaining models we have calculated the algebraic entropy as explained
in the following section.

\subsection{Growth patterns and algebraic entropy computations}
Suppose initial data is given on a line which allows the determination
of the values at all points of the lattice, for example a regular
diagonal staircase.  The multilinear relation (\ref{M}) allows us to
define evolutions in the following way: we iterate the relation by
calculating the values on diagonals moving away from the initial
staircase, as in the previous section. After cancellations we get a
sequence of degrees $d_n$.

We may in this way define four fundamental evolutions, corresponding
to initial data given on diagonals with slope $+1$ or $-1$, and
evolutions towards the four corners of the lattice. We denote them
``NE, SE, SW, NW'' by the orientation of the evolutions (towards
North-East, South-East, and so on), see Figure \ref{F:comp}.  To each
evolution we associate an entropy with the definition inspired by the
1-dimensional case~\cite{HV,BeVi99,Vi06,TrGrRa01}:
\begin{eqnarray}
\label{algent}
\epsilon_{} = \lim_{n\rightarrow\infty}\tfrac1n\;\log(d_{n}) .
\end{eqnarray}
These entropies always exist~\cite{BeVi99}, because of
the subadditivity property of the logarithm of the degree of composed
maps.

A full calculation of iterates is usually beyond reach. We can however
get explicit sequences of degrees as explained in~\cite{Vi06}.

Suppose we start from initial values distributed on a diagonal,
containing $\nu$ vertices ${V}_1, \dots,{V}_\nu$. For each of these
$\nu$ vertices, we assign to the dependent variable $x_{n,m}$ an
initial value of the form:
\begin{eqnarray}
x_{[{V}_k]} = {{\alpha_k + \beta_k \; t }\over{
\alpha_0 + \beta_0 \; t }}, \quad \; k = 1 \dots \nu
\end{eqnarray}
where $\alpha_0$, $ \beta_0$ and $ \alpha_k$, $\beta_k, (k=1..\nu)$
are arbitrary constants, and $t$ is some unknown.  We then calculate
the values of $x$ at the vertices which are within the range of our
set of initial date.  These values are rational fractions of $t$,
whose numerator and denominator are of the same degree in $t$, and that
is the degree we are looking for.

The next step is then to obtain ``degree growth'', i.e., to extract
the value of the entropy from the first few terms of the sequence
$\{d_n\}$.  One very fruitful method is to introduce the generating
function of the sequence of degrees
\begin{eqnarray}
g(s) = \sum_{k=0}^{\infty} s^k \, d_k 
\end{eqnarray}
and try to fit it with a rational fraction.  The remarkable fact
  is that this works surprisingly well, as it did for maps,
although we know that it may not always be the case~\cite{HaPr05}.
This means that we can often calculate the asymptotic behavior
measured by (\ref{algent}) from a finite beginning part of the
sequence of degrees.

We have done this calculation for all four evolutions of the 80 cases
we got.  The $4\times 80$ calculations resulted with a number of
different sequences of degrees, which could all be fitted with
rational generating functions. Some have linear growth, some are
quadratic, some have exponential growth. The sequences are given in
the following tables, containing the beginning of the sequences, the
generating function, and the numerical value $\rho$ of the growth
$\rho ^n$

Linear growth:
\begin{center}
\begin{tabular}{|c|l|c|c|}
\hline
& Sequence of degrees  & Generating function & Growth \\
\hline
 $l_1$ & $1,2,3,4,5,6,7,8,9,10,\dots$ &
$ \tfrac1{(1-s)^2}$  &  1 \\
\hline
 $l_2$ & $1,2,3,5,6,8,9,11,12,14,\dots$&
$ \tfrac{1+s+s^3}{(s+1)(1-s)^2}$ &  1 \\
\hline
\end{tabular}
\end{center}

Quadratic growth:
\begin{center}
\begin{tabular}{|c|l|c|c|}
\hline
& Sequence of degrees & Generating function & Growth \\
\hline
 {}$ q_1$ & $1, 2, 3, 5, 7, 10, 13, 17, 21, 26,\dots$&
$ \tfrac{1-s^2+s^3}{(s+1)(1-s)^3}$ & 1  \\
\hline
 {}$ q_2$ &  $1, 2, 4, 6, 9, 12, 16, 20, 25, 30,\dots$&
$ \tfrac1{(s+1)(1-s)^3}$ & 1  \\
\hline
 {}$ q_3$ &  $1, 2, 3, 5, 7, 11, 14, 20, 24, 32,\dots$&
$\tfrac{1+s-s^2+s^4+s^5}{(s+1)^2(1-s)^3}$ &  1 \\
\hline
 {}$ q_4$ &  $1, 2, 3, 5, 8, 12, 17, 23, 30, 38,\dots$&
$ \tfrac{1-s+s^3}{(1-s)^3}$ & 1  \\
\hline
 {} $q_5$ &  $1, 2, 4, 7, 11, 16, 22, 29, 37, 46,\dots$&
$  \tfrac{1-s+s^2}{(1-s)^3}$ &  1 \\
 \hline
\end{tabular}
\end{center}

Exponential growth:
\begin{center}
\begin{tabular}{|c|l|c|c|}
\hline
& Sequence of degrees & Generating function & Growth \\
\hline
{} $e_1$ &  $  1, 2, 3, 5, 8, 13, 21, 34, 55, 89,\dots$&
$ \tfrac{1+s}{1-s-s^2} $ & 1.618  \\
\hline
 {} $e_2$ &  $  1, 2, 4, 7, 12, 20, 33, 54, 88, 143,\dots$&
$\tfrac1{(1-s)(1-s-s^2)}$ & 1.618  \\
\hline
 {} $e_3$ &  $  1, 2, 3, 5, 9, 16, 28, 49, 86, 151,\dots$&
$\tfrac1{1-2 s+s^2-s^3}$ & 1.755  \\
\hline
 {} $e_4$ &  $  1, 2, 3, 5, 9, 17, 32, 60, 112, 209,\dots$&
$\tfrac{ (1-s) (s+1)}{1-2 s+s^3-s^4}$ & 1.867  \\
\hline
 {} $e_5$ &  $  1, 2, 4, 7, 13, 24, 45, 84, 157, 293,\dots$&
$\tfrac1{1-2 s+s^3-s^4}$ &  1.867 \\
\hline
 {} $e_6$ &  $  1, 2, 4, 8, 15, 28, 52, 97, 181, 338,\dots$&
$\tfrac{ (1+s) (1- s+ s^2)}{1-2 s+s^3-s^4}$ & 1.867  \\
\hline
 {} $e_{7}$ &  $  1, 2, 3, 5, 9, 17, 33, 65, 129, 257,\dots$&
$ \tfrac{1-s-s^2}{(1- 2 s)(1-s)}$ & 2.  \\
\hline
 {} $e_{8}$ &  $  1, 1, 3, 5, 11, 21, 43, 85, 171, 341,\dots$&
$ \tfrac1{(1+s)(1-2 s)}$ &  2. \\
\hline
 {} $e_{9}$ &  $  1, 1, 3, 6, 12, 24, 48, 96, 192, 384,\dots$&
$ \frac{1-s+s^2}{1-2 s}$ & 2.  \\
\hline
 {} $e_{10}$ &  $  1, 2, 4, 7, 14, 27, 54, 107, 214, 427,\dots$&
$\tfrac{1-s^2-s^3}{(1-s)(1-2 s)(1+s)}$ &  2. \\
\hline
 {} $e_{11}$ &  $  1, 2, 4, 8, 16, 32, 64, 128, 256, 512,\dots$&
$  \tfrac1{1-2 s}$ &  2. \\
\hline
 {} $e_{12}$ &  $  1, 2, 4, 8, 16, 32, 65, 133, 274, 566, \dots$&
$ \tfrac{(1+s) (1-s)^2}{1-3 s+s^2+3 s^3-2 s^4-s^6}$  & 2.067  \\
\hline
 {} $e_{13}$ &  $  1, 2, 4, 8, 16, 33, 68, 141, 292, 605, \dots$&
$ \tfrac{(1-s) (s+1)}{1-2 s-s^2+2 s^3-s^5}$ &  2.071 \\
\hline
 {} $e_{14}$ &  $  1, 2, 4, 8, 16, 33, 69, 145, 305, 642, \dots$&
$ \tfrac{1-s-s^4}{(1-s)(1-2 s-s^4)}$ & 2.107  \\
\hline
 {} $e_{15}$ &  $  1, 2, 4, 8, 17, 37, 82, 183, 410, 920,\dots$&
$ \tfrac{1-s-s^2}{(1-s)(1-2 s-s^2+s^3)}$ & 2.247  \\
\hline
 {} $e_{16}$ &  $ 1, 2, 4, 9, 20, 45, 101, 227, 510, 1146,\dots$&
$ \tfrac{(1-s) (s+1)}{1-2 s-s^2+   s^3}$ & 2.247  \\
\hline
 {} $e_{*}$ &  $ 1, 2, 4, 9, 21, 50, 120, 289, 697, 1682 \dots$&
$ \tfrac{1-s-s^2}{(1-s)(1-2 s - s^2)}$ & 2.414  \\
\hline
\end{tabular}
\end{center}

In the previous table, the value $e_*$ is given for reference, it
corresponds to the absence of factorization, and is an upper bound.
The factorization condition we have used is a rather mild one, it
says that the fourth number should be 8 or less. (The pattern $e_{16}$
does not have this cancellation, it arises in a model, which has
cancellations in two direction, but not in the other two.)

From these tables we also observe, that the sometimes a map with
asymptotically exponential growth starts with as slow growth as is
seen for linear growth, compare the patterns of
$e_1,\,e_3,\,e_4,\,e_7,\, e_8$ with $l_2$.

\subsection{Classification into parents and descendents}
We are not going to give the full listing of the 80 models obtained,
because the solution method is insensitive to subcase dependency. The
criterion was only that there is one linear factor at position $(2,2)$
and this does not yet guarantee integrability. It is possible that one
model is obtained from another one by specializing the values of the
parameters, giving a notion of descendent. This process provides a
partial ordering of the models, the descendents of a model having a
smaller entropy. It may happen that a model with non-vanishing
entropy, thus a priori non integrable, has a descendent with vanishing
entropy.

The list that is presented next is comprehensive in the sense that any
integrable case with a linear factor at $(2,2)$ will be a subcase of
one of the presented models. The list is rather short and therefore
further studies of subcases with more factorization will not be
overwhelming, but not included here.

The classification is up to transformations of the type
\[
x_{nm}\mapsto x_{nm}+a(n-n_0)+b(m-m_0)+c,
\]
where the $a,b$ terms can only be used if the result is $n,m$
independent.  We would also like to note that innocuous
reparametrizations can change the factorization properties, because
computer factorization is normally over integers. Examples of this can
be seen in Cases 2 and 5.

The ancestral models are as follows:

\subsubsection*{Case 1}
This is a homogeneous model, where one quadratic term is missing, all
others have arbitrary coefficients:
\begin{equation}
\begin{aligned} &
x_{00} x_{10} p_1 + x_{00} x_{11} p_4 + x_{10} x_{01} p_2 + x_{10}
x_{11} p_6 + x_{01} x_{11} p_3 =0
\end{aligned}
\end{equation}
The degree sequences in the four directions are $e_{15}, e_{16},
e_{16}, e_{15}$. The missing term above is $x_{00}x_{01}$, but it
could be rotated into any other side of the elementary square. Since
the model is not rotationally symmetric it can have different growth
patterns in different directions. The pattern $e_{16}$ corresponds to
growth without any cancellations at $(2,2)$, but $e_{15}$, which has
some cancellations, is obtained in the direction of the test. As for
its subcases, we just mention that if $p_3=0$, i.e., the opposite side
$x_{10}x_{11}$ is also missing, then the map is linearizable.

\subsubsection*{Case 2}
This is also homogeneous, but it has ``symmetric cross'' $x_{00}
x_{11} + x_{10} x_{01}$ and all other coefficients are arbitrary
\begin{equation}
x_{00} x_{10} p_1 + x_{00} x_{01} p_5(p_1p_3+p_2) + (x_{00} x_{11} +
x_{10} x_{01})p_2 + x_{10} x_{11} p_6  + x_{01} x_{11} p_3(p_5p_6-p_2)=0
\end{equation}
This has quadratic growth $q_5$ is all directions. This result
illustrated the role of parametrization in factorization: there are
several ways to reparametrize $p_3,p_5$ so that combinations
$p_5(p_1p_3+p_2)$ and $p_3(p_5p_6-p_2)$ look simpler, but this could
easily lead to one quadratic rather than two linear factors at the
corner.

\vskip 1cm
The next four models have a free nonhomogeneous constant term.

\subsubsection*{Case 3}
Here there are completely arbitrary linear and constant terms, while
in the quadratic part two opposite sides are missing ($x_{00}x_{10}$
and $x_{01}x_{11}$)
\begin{equation}\begin{aligned} &
x_{00} x_{01} p_5 + x_{00} x_{11} p_4 + x_{10} x_{01} p_2 + x_{10}
x_{11} p_6
+x_{00} r_1 + x_{10} r_2 + x_{01} r_3 + x_{11} r_4+u=0
\end{aligned}
\end{equation}
The growth pattern is  $e_{11}$ in all directions, so as it stands
the model is not integrable. Among its subcases we would like to
mention
\begin{equation}
(p_3 x_{01} + x_{00}) (p_3 x_{11} + x_{10})+(p_3 x_{01} + x_{10})r_3=0
\end{equation}
and
\begin{equation}\label{q3model} (x_{00} - x_{01}) (x_{10} - x_{11}) +
(x_{00} - x_{11}) r_4 + (x_{01} - x_{10}) r_3+u=0
\end{equation}
both integrable, with growth patterns $q_3$ in all directions.

\subsubsection*{Case 4}
Here is the homogeneous part the cross is missing, and other terms
have arbitrary coefficients, except that there is a relation between
the $r_i$ coefficients:
\begin{equation}
\begin{aligned}
 &  x_{11} x_{10} p_6
 + x_{11} x_{01} p_3
 + x_{10} x_{00} p_1
 + x_{01} x_{00} p_5
\\& + x_{11} p_3 p_6 r_4
 + x_{10} p_6 r_2
 + x_{01} p_3 r_3
 + x_{00} ( - p_1 p_5 r_4 + p_1 r_3 + p_5 r_2)
 + u =0.
\end{aligned}
\end{equation}
The parameters have been scaled to simplify the model, but still nicer
forms of writing it may exist. The growth pattern is again $e_{11}$ in
all directions. This contains as subcases the following
integrable ones:
\begin{equation}
(x_{00} x_{01} + x_{10} x_{11}) p_5 + (x_{00} x_{10} + x_{01}
x_{11})p_3 +u=0,
\end{equation}
\begin{equation}
(x_{00} - x_{11}) (x_{01} - x_{10})+r_1 (x_{00} + x_{10} + x_{01} + x_{11})+u=0,
\end{equation}
which include H1,H2,H3 of \cite{ABS}

\subsubsection*{Case 5}
The following model is integrable, with pattern  $q_5$ is all directions:
\begin{equation}
(x_{11}+x_{00})(x_{10}+x_{01})p_2
+(x_{11}+x_{01})(x_{00}+x_{10})p_3+p_3p_2(p_3+p_2)u^2=0,
\end{equation}
or in another way of writing
\begin{equation}
(x_{ 1 1}x_{ 1 0} + x_{0 1}x_{0 0})(p_2 + p_3)
 + (x_{ 1 1}x_{0 1} + x_{ 1 0}x_{0 0})p_3
 + (x_{ 1 1}x_{0 0} + x_{ 1 0}x_{0 1})p_2
 + p_3p_2(p_3+p_2)u^2=0.
\end{equation}
The above parameterization of the constant term is necessary in order
to have two linear factors at $(2,2)$, otherwise we would get one
quadratic factor.  This case becomes Q1 with scaling $x_{nm}\mapsto
(-1)^{n-n_0} x_{nm}$ (it is not separately listed in \cite{ABS}, but the
situation is similar to A1).

\subsubsection*{Case 6}
\begin{equation}\begin{aligned} &
 x_{11} x_{00}+ x_{10} x_{01}
+(x_{11} x_{01}+ x_{10} x_{00})p_3
-(x_{11} x_{10}+ x_{01} x_{00})(p_3+1)\\
&+(x_{11} - x_{00}) r_4 + (x_{10} - x_{01}) r_2 -
(u (p_3+1)+r_4)(u p_3+r_4)+u r_2=0,
\end{aligned}
\end{equation}
which is integrable with growth $q_5$ in all directions. If
$r_4=r_2=0$ we get Q1 of \cite{ABS}. In fact, if $p_3\neq 0,-1$ we can
apply an $n,m$ dependent translation to put $r_2=r_4=0$.

\vskip 1cm
The remaining maps have linear and constant terms, but the constant
term depends on the linear terms. There are also interesting relationships
between the parameters of the homogeneous terms. There may be other
interesting representations obtainable with the transformation
$x_{ij}\to\mu x_{ij}+\nu$.

\subsubsection*{Case 7}
\begin{equation}
\begin{aligned} &
   x_{11} x_{10} p_3^2
 + x_{01} x_{00} p_6^2
 + x_{11} x_{01} p_3p_1
 + x_{10} x_{00} p_3p_1^{-1} p_6^2 \\&
 + (x_{11} p_3
 + x_{10} p_3
 + x_{01} p_6
 + x_{00} p_6) r_1
 + r_1^2=0
\end{aligned}
\end{equation}
with growth $e_{11}$ in all directions.

\subsubsection*{Case 8}
\begin{equation}
\begin{aligned} &
   x_{11} x_{10} p_6^2
 + x_{01} x_{00} p_3^2
 + x_{11} x_{01} p_1^{-1} p_6 (p_3 - 1)
 + x_{10} x_{00} p_1 p_6 (p_3 - 1)
 + (x_{11} x_{00}
 + x_{10} x_{01}) p_6  \\ &
 + (x_{11} p_6
 + x_{10} p_6
 + x_{01} p_3
 + x_{00} p_3) r_4
 + r_4^2=0
\end{aligned}
\end{equation}
also with growth $e_{11}$ in all directions.

\subsubsection*{Case 9}
\begin{equation}
\begin{aligned} &
( x_{11}x_{00} + x_{10}x_{01})
+ p_3 (x_{11}x_{01} + x_{10}x_{00})
 + p_6 x_{11}x_{10}
 +x_{01}x_{00}p_6^{-1}(p_3 - 1)^2\\ &
+ r_3(p_6-p_3+1) (x_{01} + x_{00})
+p_6(p_6+1)r_3^2=0.
\end{aligned}
\end{equation}
This has growth $e_{14}$ in all directions. In the subcase $p_6=p_3-1$
it becomes the integrable case A1 of \cite{ABS}, on the other hand, if 
$p_6\neq p_3-1$ it can be translated into the more symmetric form
\begin{equation}
\begin{aligned} &
( x_{11}x_{00} + x_{10}x_{01})
+ p_3 (x_{11}x_{01} + x_{10}x_{00})
 + p_6 x_{11}x_{10}
 +x_{01}x_{00}p_6^{-1}(p_3 - 1)^2\\ &
+ r_4 (x_{11} + x_{10} + x_{01} + x_{00})
+r_4^2=0.
\end{aligned}
\end{equation}

\subsubsection*{Case 10}
\begin{equation}\begin{aligned} &
(x_{00} x_{10} + x_{01} x_{11}) + (x_{00} x_{11} + x_{01} x_{10})p_3
  + (p_3 - 1) (x_{00} x_{01} + x_{10} x_{11})\\
&(x_{00}-x_{01}+x_{10}-x_{11})r_4+r_4^2=0,
\end{aligned}
\end{equation}
also with  growth $e_{14}$ in all directions.

\subsubsection*{Case 11}
\begin{equation}
\begin{aligned}
&(x_{00} x_{11} + x_{01} x_{10}-x_{01}x_{00}) p_4 + p_1 x_{00} x_{10}
+(x_{00}p_1+x_{10}+x_{11}p_4)r_4+r_4^2=0
\end{aligned}
\end{equation}  
with growths $e_5 , e_{16} , e_{16} , e_6$ in the for directions.

\subsubsection*{Case 12}
\begin{equation}\begin{aligned} &
x_{00} x_{10} p_1 + x_{01} x_{11} p_3  + x_{00} x_{01} (p_1 p_3 - 1) + x_{00} x_{11} +
x_{10} x_{01}\\
&+(x_{00} p_1 + x_{10} + x_{01} p_3 + x_{11}) r_4+r_4^2=0
\end{aligned}
\end{equation}
with growths $e_5 , e_{16} , e_{16} , e_{16}$.

The known models referred to in section (\ref{known_models}), that have
linear factors, all appear in our analysis.

\section{Some integrable cases}
\label{S:skkn}
Due to the search condition (one linear factor at $(2,2)$) the results
are not comprehensive as far as integrable cases are concerned. We can
say that they  must all be subcases of the cases enumerated above.

Case 2 is integrable as it stands. We can also write it as 
\begin{equation}
x_{00} x_{10} c_1 + x_{00} x_{01} c_5 + (x_{00} x_{11} +
x_{10} x_{01})c_2 + x_{10} x_{11} c_6  + x_{01} x_{11} c_3=0
\end{equation}
and the statement is that this lattice map is integrable for all
values of the five parameters $c_i$. If we next consider integrability
in the sense of consistency, then the parameters $c_i$ must have
specific forms in terms of the spectral parameters associated with the
lattice directions. The result is
\begin{equation}
(x x_{[ij]}+x_{[i]}x_{[j]})(\alpha_i-\alpha_j)
+x x_{[i]}(\beta_j+\alpha_j^2)/\gamma_j
-x x_{[j]}(\beta_i+\alpha_i^2)/\gamma_i
-x_{[i]}x_{[ij]}\gamma_i
+x_{[j]}x_{[ij]}\gamma_j=0
\end{equation}
where the parameters associated with the third direction are restricted
by
\begin{equation}
\left|\begin{matrix}
1&\alpha_1&\beta_1\\
1&\alpha_2&\beta_2\\
1&\alpha_3&\beta_3\\
\end{matrix}\right|=0.
\end{equation}
An immediate way to resolve this constraint is to take 
\[
\alpha_i=\mu\, p_i^2+\nu,\quad\beta_i=\rho\, p_i^2+\sigma.
\]
The various homogeneous KdV equations are obtained as subcases (with
{\em one} spectral parameter), the modified KdV of \eqref{mkdv} is
obtained with $\gamma_i=p_i,\, \rho=-1,\,\mu=\nu=\sigma=0$ and the
Schwarzian KdV of \eqref{skdv} with $\gamma_i=-p_i^2,\,
\mu=-1,\,\rho=\nu=\sigma=0$.  Furthermore the generalized form
of~\cite{NiQuCa83} is included with $\gamma_i=-(p_i+a)(p_i+b),
\,\mu=-1,\,\nu=0, \,\rho=-(a^2+b^2), \,\sigma=a^2b^2$.  By suitable
scaling of type $x_{nm}\mapsto A^{n-n_0}B^{m-m_0}\,x_{nm}$ one can
reduce this model to Q3 with $\delta=0$.

It is perhaps worth mentioning that if some of the $c_i$ vanish we may
get asymmetric models with different growth in different directions,
for example
\begin{equation}
x_{00} x_{10} c_1 + x_{00} x_{01} c_5 + (x_{00} x_{11} +
x_{10} x_{01})c_2=0,
\end{equation}
has growth patterns $q_2,q_5,q_1,q_5$, and
\begin{equation}
x_{00} x_{10} c_1 + x_{10} x_{11} c_6  + x_{01} x_{11} c_3=0
\end{equation}
grows as  $q_5,q_4,q_4,q_5$. It is not clear how such special cases
carry over to the consistency approach.

As an interesting model worth further study we would like to present
\eqref{q3model}
\[
(x_{00}-x_{01})(x_{10}-x_{11})+(x_{00}-x_{11})r_4+(x_{01}-x_{10})r_3+u=0.
\]
It is integrable with growth patters $q_3$ in all directions. But this
model is not symmetric and therefore its consistency formulations is
problematic: How should the maps on the other sides of the cube be
oriented, or do we perhaps need entirely different maps there?

\section{Conclusions}
We have analyzed the factorization process for a class of
two-dimensional lattice models (``quad models''). By solving the
resulting equations we obtained a list of models. We then conducted an
entropy analysis on the results. This provided an ordering among the
models, by increasing complexity.

This list of minimally factoring models contains both models with
vanishing entropy, and models with different non-vanishing entropy.
These models with  non-vanishing entropy may contain integrable models
as subcases, but this has not been comprehensively analyzed here.

More can be done along the lines we have followed: one could insist on
stronger factorization requirements, or one could start from the most
general quartic defining relation, rather than restricting on
quadratic ones as we did here. Both of these is beyond the scope of
this paper.

{\bf Acknowledgments.} C Viallet acknowledges support from European Union
network ENIGMA, MRTN  CT~2004~56~52, and wishes to thank University of Turku,
Dept of Physics for hospitality.

\end{document}